\documentclass{article}
\usepackage{spconf,amsmath,graphicx}
\usepackage{booktabs}
\usepackage{xspace}
\usepackage{hyperref}
\usepackage[utf8x]{inputenc}

\title{MOCKINGJAY: UNSUPERVISED SPEECH REPRESENTATION LEARNING WITH DEEP BIDIRECTIONAL TRANSFORMER ENCODERS}
\name{Andy T. Liu \qquad Shu-wen Yang \qquad Po-Han Chi \qquad Po-chun Hsu \qquad Hung-yi Lee}
\address{National Taiwan University\\
         College of Electrical Engineering and Computer Science\\
         \{r07942089, r08944041, r08942074, r07942095, hungyilee\}@ntu.edu.tw}
\begin{document}
%
\maketitle
\begin{abstract}
We present Mockingjay as a new speech representation learning approach, where bidirectional Transformer encoders are pre-trained on a large amount of unlabeled speech.
Previous speech representation methods learn through conditioning on past frames and predicting information about future frames. Whereas Mockingjay is designed to predict the current frame through jointly conditioning on both past and future contexts. 
The Mockingjay representation improves performance for a wide range of downstream tasks, including phoneme classification, speaker recognition, and sentiment classification on spoken content, while outperforming other approaches. 
Mockingjay is empirically powerful and can be fine-tuned with downstream models, with only 2 epochs we further improve performance dramatically.
In a low resource setting with only 0.1\% of labeled data, we outperform the result of Mel-features that uses all 100\% labeled data.  

\end{abstract}


\section{Introduction}
\label{sec:intro}

\begin{figure}[t]
  \centering
  \includegraphics[width=\linewidth]{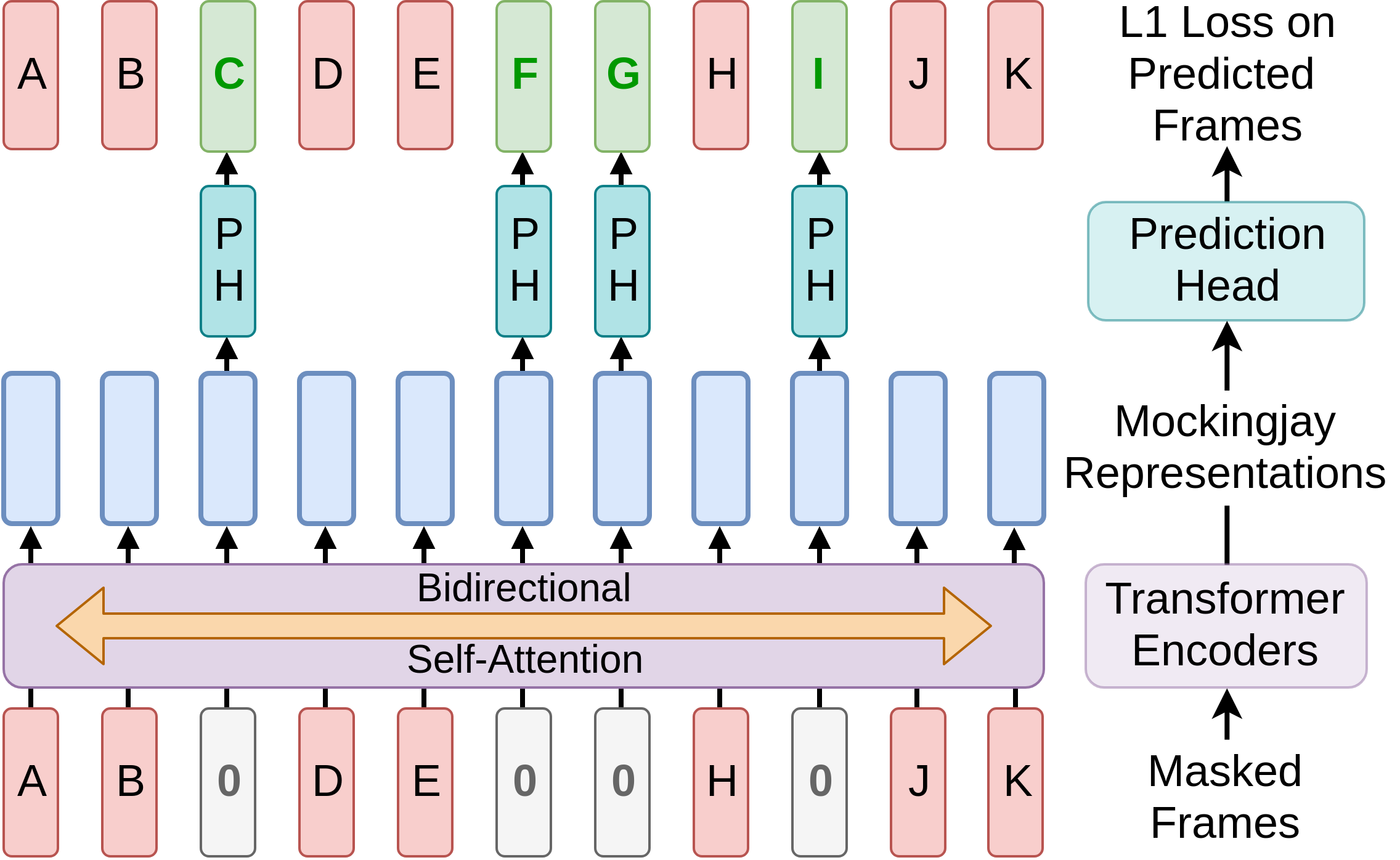}
  \caption{The proposed Masked Acoustic Model pre-training task, 15\% of input the frames are masked to zero at random.}
  \label{fig:training}
  \vspace{-10pt}
\end{figure}

The goal of speech representation learning is to find a transform from speech that makes high-level information more accessible to SLP (Speech and Language Processing) downstream tasks, as speech signal possess a rich set of acoustic and linguistic content, including phonemes, words, semantic meanings, tone, speaker characteristics, and even sentiment information. 
In this paper, we propose Mockingjay to learn speech representations through unsupervised training without the use of any labels.
We use multi-layer transformer encoders and multi-head self-attention \cite{TRANSFORMER} to achieve bidirectional encoding; this framework allows our model to consider past and future contexts at the same time. 
To achieve unsupervised pre-training for speech representations, Mockingjay learns under the proposed Masked Acoustic Model (MAM) task. During training, masked frames are given, and the model learns to reconstruct and predict the original frames.
Hence we gave the name Mockingjay, a bird that mimics sound.
The proposed framework is illustrated in Figure~\ref{fig:training}.

\subsection{Related work}
\label{sssec:related work}

Unsupervised speech representation learning \cite{WAVENET_AUTOENCODERS, SPEECH2VEC, AUDIO_WORD2VEC, CPC, APC, WAV2VEC, VQWAV2VEC, ROBUSTCPC, MBV} is effective in extracting high-level properties from speech. SLP downstream tasks can be improved through speech representations because surface features such as log Mel-spectrograms or waveform can poorly reveal the abundant information within speech.

Contrastive Predictive Coding (CPC)~\cite{CPC} and wav2vec~\cite{WAV2VEC} use a multi-layer CNN to encode past context, representations are learned by predicting the future in latent space under a contrastive binary classification task. 
Autoregressive Predictive Coding (APC)~\cite{APC} uses autoregressive models to encode temporal information of past acoustic sequence; the model predicts future frames like an RNN-based language model~\cite{RNN}, optimized with reconstruction loss. 
Unidirectional models are commonly used in the previous approaches \cite{WAVENET_AUTOENCODERS, SPEECH2VEC, AUDIO_WORD2VEC, CPC, APC, WAV2VEC}. However, this constraint on model architectures limits the potential of speech representation learning.

\begin{figure}[t]
  \centering
  \includegraphics[width=\linewidth]{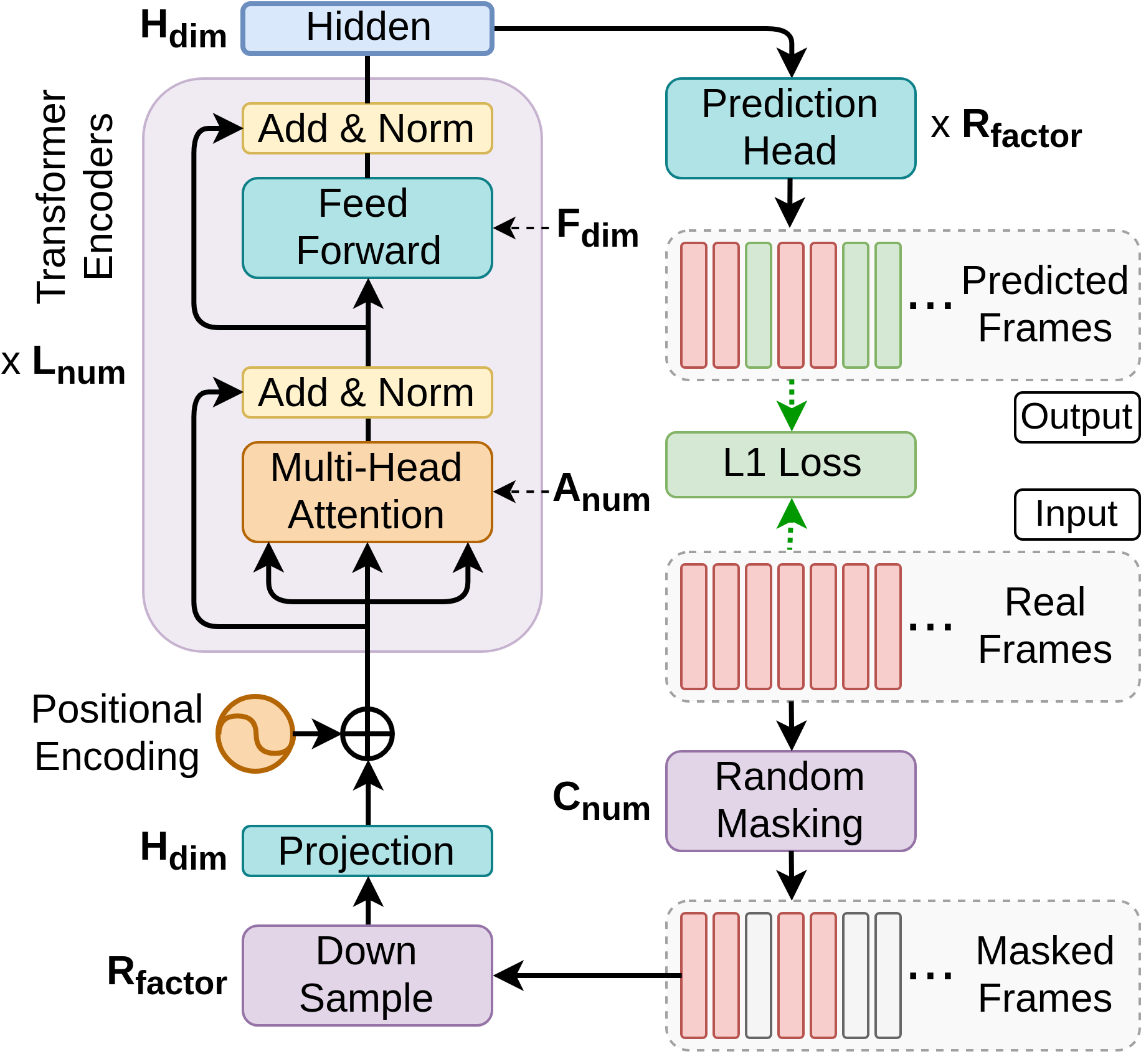}
  \caption{The proposed Mockingjay training framework.}
  \label{fig:model}
  \vspace{-10pt}
\end{figure}

The recently proposed vq-wav2vec~\cite{VQWAV2VEC} approach attempts to apply the well-performing Natural Language Processing (NLP) algorithm BERT~\cite{BERT} on continuous speech. Input speech is discretized to a K-way quantized embedding space, so continuous speech could act like discrete units similar to word tokens in NLP tasks. In vq-wav2vec~\cite{VQWAV2VEC}, an exhaustive two-stage training pipeline with massive computing resources are required to adapt speech to NLP algorithm, as the quantization process is against the continuous nature of speech. Unlike \cite{VQWAV2VEC} that adapts speech to BERT~\cite{BERT} through quantization, the proposed approach can be seen as a modified version of BERT~\cite{BERT} for direct application on continuous speech.


\subsection{Proposed Method}
\label{sssec:proposed method}
Unlike previous left-to-right unidirectional approaches that only consider past sequences to predict information about future frames, the proposed method allows us to train a bidirectional speech representation model, alleviating the unidirectionality constraint of previous methods.
As a result, the Mockingjay model obtains substantial improvements in several SLP tasks.
Moreover, as previous approaches restrict the power of the pre-trained models to representation extraction only \cite{CPC, APC, WAV2VEC, VQWAV2VEC}, the proposed method is robust as it can be fine-tuned easily on downstream tasks. We show that fine-tuning for 2 epochs easily acquires significant improvement. 

The proposed approach outperforms other representations and features. When compared to the commonly used log Mel-features, we outperformed it by 35.2\% (absolute improvement) for phoneme classification accuracy, 28.0\% (absolute improvement) for speaker recognition accuracy, and 6.4\% (absolute improvement) for sentiment discrimination accuracy on a spoken content dataset unseen during pre-train.
We also experiment in low resource settings to show that Mockingjay is capable of improving supervised training in real-life low-resource scenarios.
With only 0.36 hours (0.1\%) of transcribed speech, the proposed approach outperforms Mel-features with 360 hours (100\%) of labels. 

\section{MOCKINGJAY}
\label{sec:mockingjay}
In this section, we first introduce model architecture and its designs, secondly we explain the proposed unsupervised context prediction task, and finally we explain how the proposed model is used with downstream task models.

\subsection{Model Architecture}
\label{sssec:model architecture}
We use a multi-layer Transformer encoder with multi-head self-attention for left-and-right bidirectional encoding, this architecture is illustrated in Figure~\ref{fig:model}. 
Each encoder layer has two sub-layers, the first is a multi-head self-attention network, and the second is a feed-forward layer, each sub-layer has a residual connection followed by layer normalization \cite{LAYERNORM}, based on the design described in \cite{TRANSFORMER}.
All encoder layers in the model, as well as the sub-layers, produce outputs of identical dimensions denoted as $H_{dim}$.  
In Figure~\ref{fig:model}, we denote the feed-forward size as $F_{dim}$, the number of self-attention heads as $A_{num}$, and the total of Transformer layers as $L_{num}$. The Mockingjay representations can be extracted from the Transformer encoders' hidden state and labeled as $Hidden$, we explain how they are used as representations in Section~\ref{sssec:incorporating with downstream tasks}.

Since Transformer encoders contain no recurrence and convolution, we use positional encoding to make our model aware of the input sequence order \cite{TRANSFORMER}. As direct addition of acoustic features to positional encoding may lead to potential training failure \cite{DOWNSAMPLE}, the input frames are first projected linearly to the hidden dimension of $H_{dim}$ before adding with positional encoding \cite{TRANSFORMER_ASR}. We use sinusoidal positional encoding instead of learnable positional embeddings~\cite{LEARNABLE_EMBEDDING} because acoustic features can be arbitrarily long with high variance \cite{TRANSFORMER_ASR}.
We apply downsampling on input features to adapt our model to long sequences. To reduce the length of frames by a factor of $R_{factor}$, we use the reshape technique from \cite{DOWNSAMPLE, TRANSFORMER_ASR} by stacking $R_{factor}$ consecutive frames into one step.

\subsection{Masked Acoustic Modeling} 
\label{sssec:masked acoustic modeling}

We propose the Masked Acoustic Modeling task, where we randomly select 15\% of the input frames, and the model predicts the selected frames based on its left and right context, as illustrated in Figure~\ref{fig:training}. 
During training, we add a prediction head consists of two layers of feed-forward network with layer-normalization, using the last encoder layer as it's input. We use L1 Loss to minimize reconstruction error between prediction and ground-truth frames on the selected 15\%.
The prediction head is not used once the model is trained. 

During training, for the selected 15\% of frames, 1) we mask it all to zero 80\% of the time, 2) replace all with a random frame 10\% of the time, and 3) leave the frames untouched 10\% of the time.\footnote{This process is similar to the Cloze task~\cite{MLM} and the Masked Language Model task from BERT~\cite{BERT}, but we mask frames of speech to zero instead of using the MASK token.}
We introduce this sub-random process instead of always masking the frames to alleviate the mismatch between training and inference, as masked frames do not appear during inference time. 
Note that in contrast to BERT~\cite{BERT}, where the sub-random process is performed token-wise on the i-th chosen token, our sub-random process is performed utterance-wise. In other words, our model may receive inputs as ground-truth frames for 3) 10\% of the time, rather with some of the inputs always augmented as in \cite{BERT}.

To avoid the model exploiting local smoothness of acoustic frames, we propose additional consecutive masking where we mask consecutive frames $C_{num}$ to zero. The model is required to infer on global structure rather than local information.
We also use dynamic masking~\cite{ROBERTA} where masking patterns are sampled from an uniform distribution every time we feed a sequence to the model, unlike the static mask employed in \cite{BERT} where masking is performed during data preprocessing.
We only use a single context prediction task to train our representation model, as suggested by \cite{ROBERTA}. Unlike BERT~\cite{BERT} and ALBERT~\cite{ALBERT} that needs two tasks to train their language model.
In our preliminary experiments, we found that the sentence prediction task used in \cite{BERT, ALBERT} is not helpful, as additional tasks can potentially harm training behavior. We do not include details due to space limitations. 



\subsection{Incorporating with Downstream Tasks}
\label{sssec:incorporating with downstream tasks}
Mockingjay representations are essentially the Transformer encoder hidden states.
There are many ways to incorporate learned representations to downstream tasks. 
In this work, we mainly extract representations from the last layer. 
However, we also expose the deep internals of Mockingjay to downstream models, where we use a mixture of representations from all layers, similar to the ELMO~\cite{ELMO} approach. In other words, we use a learnable weighted sum to integrate hidden states from all layers.
Last but not least, a pre-trained Mockingjay model can be fine-tuned with downstream models to create improving results, we update the pre-trained Mockingjay together with random initialized downstream task models.

\section{IMPLEMENTATION}
\label{sec:implementation}
In this work, we use two types of features as our model's output reconstruction target: the Mel-scale spectrogram and the linear-scale spectrogram.
As Mel-scale spectrogram is a more concise acoustic feature when compared to linear-scale spectrogram, we propose two model settings: \textit{BASE} and \textit{LARGE}. 
Both of these models take Mel-features as input, and transform input Mel-features into high-level representations.
They use the same hidden dimension size of $H_{dim}$=768, feed-forward size of $F_{dim}$=3072, and attention heads of $A_{num}$=12, with the exception of layer number $L_{num}$, downsample factor $R_{factor}$, and consecutive masking number $C_{num}$, the differences in model settings are listed in Table~\ref{tb:model_settings}. We further analyze their differences in our experiment section.

\begin{table}[th]
\vspace{-5pt}
\caption{The proposed BASE and LARGE model settings}
  \label{tb:model_settings}
  \centering
\begin{tabular}{c | cc }
    Model & BASE & LARGE \\
    \toprule
    Target & Mel & Linear \\
    $L_{num}$ & 3 & 12 \\
    $R_{factor}$ & 1 & 3 \\
    $C_{num}$ & 7 & 3 \\
    parameters & 21.4M & 85.4M \\
    \bottomrule
\end{tabular}
\vspace{-10pt}
\end{table}

The proposed Mockingjay models are pre-trained on the LibriSpeech \cite{LIBRISPEECH} corpus train-clean-360 subset.
We use Adam \cite{ADAM}
where learning rate is warmed up over the first 7\% of 500k total training steps to a peak value of 4e-4 and then linearly decayed. A dropout~\cite{DROPOUT} of 0.1 is applied on all layers and attention weights. For downstream task fine-tuning, most of the hyperparameters are the same as in pre-training, with the exception of a learning rate of 4e-3, and the number of training epochs is set to 2 (which is approximately 50k steps). We train with a batch size of 6 using a single 1080Ti GPU.  
We provide pre-trained models with our implementation, they are publicly available for reproducibility\footnote{https://github.com/andi611/Mockingjay-Speech-Representation}.

\section{EXPERIMENT}
\label{sec:experiment}
Following previous works \cite{WAVENET_AUTOENCODERS, SPEECH2VEC, AUDIO_WORD2VEC, CPC, APC, WAV2VEC, VQWAV2VEC}, we evaluate different features and representations on downstream tasks, including: phoneme classification, speaker recognition, and sentiment classification on spoken content.
For a fair comparison, each downstream task uses an identical model architecture and hyperparameters despite different input features. 

We report results from 5 of our settings:
1) \textit{BASE} and 2) \textit{LARGE} where Mockingjay representations are extracted from the last encoder layer, 3) the \textit{BASE-FT2} where we fine-tune \textit{BASE} with random initialized downstream models for 2 epochs, and 4) the \textit{BASE-FT500} where we fine-tune for 500k steps, and finally 5) the \textit{LARGE-WS} where we incorporate hidden states from all encoder layers of the \textit{LARGE} model through a learnable weighted sum.
We did not fine-tune the \textit{LARGE} model, as it is meant for extracting representations.
Empirically we found that even with supervised training, a random initialized Mockingjay model followed by any downstream model is hard to be trained from scratch. This shows that the proposed pre-training is essentially indispensable.

\vspace{-10pt}
\subsection{Comparing with other representations}
\label{sssec:other representations}
The proposed approaches are mainly compared with APC~\cite{APC} representations, as they also experiment on phone classification and speaker verification. As reported in \cite{APC}, the APC approach outperformed CPC representations \cite{CPC, WAV2VEC, ROBUSTCPC} in both two tasks, which makes APC suitable as a strong baseline. 
APC uses an unidirectional autoregressive model.
We compare the proposed approach with APC to show that our bidirectional approach has advantages in speech representation learning.
For fair comparison, we pre-train APC using their official implementations with the reported ideal parameters and settings, but expand the model's hidden size to $H_{dim}$=768 to match ours. We also report results on 160-dimensional log Mel-features, which helps evaluate the accessibility of speech information from regular acoustic features.
\vspace{-10pt}

\begin{figure}[t]
  \centering
  \includegraphics[width=\linewidth]{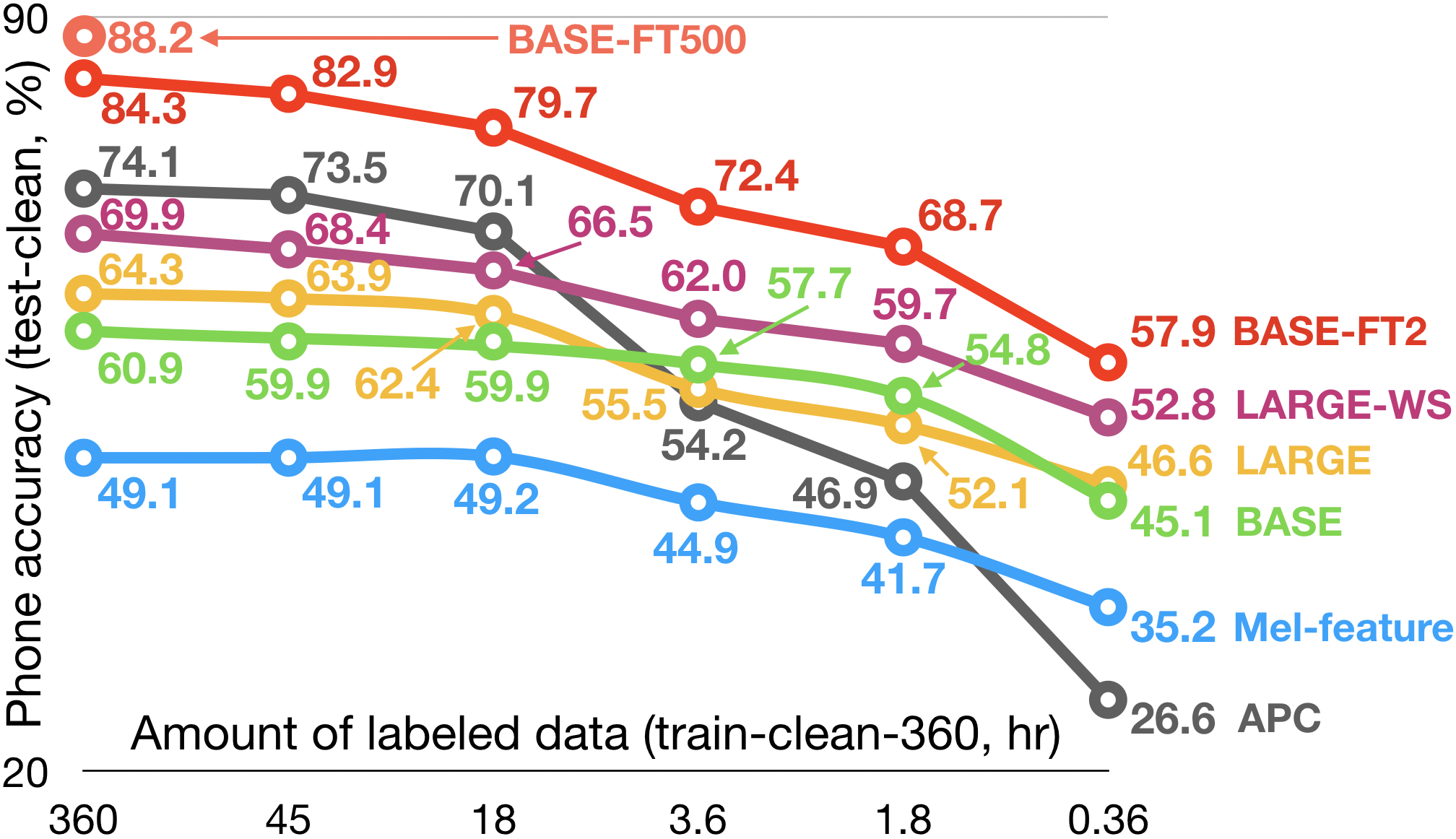}
  \caption{Comparing representations with phone classification accuracy across different amount of transcribed data.}
  \label{fig:phone}
  \vspace{-10pt}
\end{figure}

\subsection{Phoneme Classification}
\label{sssec:phoneme classification}
To measure the accessibility of phonetic information, we train linear phone classifiers using Mel-features, APC and Mockingjay representations from the LibriSpeech train-clean-360 subset.
We obtain force-aligned phoneme sequences with the Montreal Forced Aligner~\cite{ALIGNMENT}, where there are 72 possible phone classes. 
Testing results on the LibriSpeech test-clean subset are presented in Figure~\ref{fig:phone}.
In the case where all 360 hours of labels are used to train the classifier, \textit{BASE} and \textit{LARGE} representations increase 11.8\% and 15.2\% accuracy from Mel-features.
The \textit{BASE-FT2} model outperforms all representations after 2 epochs of fine-tuning, with 10.2\% and 35.2\% absolute improvement over APC and Mel-features, respectively. We observe that fine-tuning for 2 epochs is enough to reveal our method's potential as there is only a small gap (3.9\%) between \textit{BASE-FT2} and \textit{BASE-FT500}. Furthermore, \textit{LARGE-WS} improves over \textit{LARGE}, just as we expected.

To demonstrate how pre-training on speech can improve supervised training in resource constrained scenarios where human labels are short, we train the classifier with reduced amount of training data. 
The performance variation of different methods are plotted in Figure~\ref{fig:phone}, where we measure over various intervals of constrained training data to observe performance drop.  
Both \textit{BASE} and \textit{LARGE} increase accuracy over Mel-features across various amount of transcribed data. Whereas the APC approach performs well on the full resource but fails to generalize for limited amount of labeled data. In the case where there are only 0.36 hours of data available, we improve accuracy by 22.7\% (an absolute improvement from Mel-features). Note that with only 0.36 hours (0.1\%) of labels available, \textit{BASE-FT2} (57.9\% acc) even outperformed Mel-features (49.1\% acc) that uses all 360 hours (100\%) of labeled data. We conclude that pre-training Mockingjay on speech substantially improves the performance on supervised task that requires human annotations. 

\vspace{-8pt}
\subsection{Speaker Recognition}
\label{sssec:speaker recognition}
To demonstrate that the proposed approach performs constantly for all SLP downstream tasks, we report speaker recognition results on the LibriSpeech 100 hour selected subset, where train/test split is performed randomly with a 9:1 ratio, and there are 63 possible speakers.
We trained a simple one-layer RNN classifier for speaker recognition using different representations, results are listed in Table~\ref{tb:result} for comparison.
The proposed \textit{BASE} and \textit{LARGE} representations outperformed both APC and Mel-Features. \textit{BASE-FT2} further improves upon \textit{BASE} while achieving the highest accuracy, whereas \textit{LARGE-WS} also outperforms \textit{LARGE}.

\vspace{-8pt}
\subsection{Sentiment Classification on Spoken Content}
\label{sssec:sentiment classification}
To demonstrate domain invariant transferability of the proposed representation across different datasets, the Mockingjay model is pre-trained on LibriSpeech and applied on the MOSEI~\cite{MOSEI} dataset. We also use a simple one-layer RNN classifier, where the model is trained to extract linguistic meanings from speech and discriminates between sentiments. 
The results listed in Table~\ref{tb:result} lead to an identical conclusion as in the speaker recognition task discussed above. Except that in the case of sentiment classification, \textit{LARGE-WS} achieved the highest score without the need of fine-tuning, demonstrating that a deeper model has great potential for extracting general speech representations. To conclude this section, we claim that the proposed representations are general and can be used on datasets with various unseen domains.
\vspace{-5pt}

\begin{table}[t]
\vspace{-10pt}
\caption{Comparing different methods with different tasks.}
  \label{tb:result}
  \centering
\begin{tabular}{l|c|c}
    \toprule
    \textbf{Methods} & \textbf{Speaker (acc)} & \textbf{Sentiment (acc)} \\
    \midrule
    Mel-Features        & 70.06 & 64.63  \\
    APC                 & 85.88 & 65.97 \\
    Base                & 94.54 & 67.38 \\
    BaseFT2             & \textbf{98.05} & 68.45 \\
    Large               & 96.26 & 70.07\\
    LargeWS             & 96.40 & \textbf{71.05} \\
    \bottomrule
\end{tabular}
\vspace{-10pt}
\end{table}

\section{CONCLUSION}
\label{sec:conclusion}
The proposed representation contains a variety of knowledge, including but not limited to phonetic, speaker, and sentiment information. We improve performance for a wide range of downstream tasks, and show promising results in low resource settings, as the learned speech representations are robust and can be transferred to different tasks across different datasets. In future work, we will investigate and deploy Mockingjay representations on more downstream SLP tasks, including ASR, voice conversion, and speech translation.

\vfill
\pagebreak
\bibliographystyle{IEEEbib}
\bibliography{strings,refs}

\end{document}